# Guiding-centre Lagrangian and quasi-symmetry

Ted Jacobson[1]

[1] Maryland Center for Fundamental Physics, University of Maryland, College Park, MD 20742, USA

**Corresponding author:** Ted Jacobson, jacobson@umd.edu

A charged particle in a suitably strong magnetic field spirals along the field lines while slowly drifting transversely. This note provides a brief derivation of an effective Lagrangian formulation for the guiding-centre approximation that captures this dynamics without resolving the gyro motion. It also explains how the effective Lagrangian may, for special magnetic fields, admit a 'quasi-symmetry' which can give rise to a conserved quantity helpful for plasma confinement in fields lacking a geometric isometry. The aim of this note is to offer a pedagogical introduction and some perspectives on this well-established subject.

---

## 1. Introduction

The dynamics of a charged particle in a magnetic field that is nearly constant on the length and time scales of the particle's gyro motion has a well-known approximate description in which only the so-called 'guiding centre' of the motion is tracked, while the gyro motion is unresolved. The guiding-centre motion mostly follows the magnetic field lines, but includes a perpendicular velocity component, the so-called drift velocity, which arises from gradients in magnetic field strength and direction as well as from an electric field or other potential energy gradient. The guiding-centre approximation is a longstanding tool in plasma physics (Alfvén 1950; Northrop 1963; Hazeltine & Waelbroeck 1998).

The aims of this note are to (i) provide a simple, intuitive derivation of an effective Lagrangian that governs the guiding-centre motion, (ii) show how the corresponding Euler–Lagrange equations produce both the parallel acceleration and perpendicular drift velocity equations, (iii) explain how the notion of quasi-symmetry – which produces an approximately conserved quantity useful for improved plasma confinement in stellarators – is nothing but the symmetry of the background fields appearing in the Lagrangian and (iv) illustrate with an example found in the literature. These results have all been obtained before by other researchers. If anything is original in this note it is only the forms in which the ideas, derivations and equations are presented. A technical difference from most presentations is that indexed tensor methods, as well as exterior calculus, are employed, instead of the commonly used vector calculus. To keep the logical flow of ideas in the forefront, a large number footnotes are included, mostly adding explanation of what may be unfamiliar mathematical techniques. Perhaps this concise presentation can serve as a useful entry point to the subject for other newcomers like myself, or as an illustration of these mathematical techniques for veterans of the subject.

## 2. Effective Lagrangian

The Lagrangian formulation of guiding-centre motion seems to have originated with Taylor (1964). Littlejohn (1983) later derived the effective Lagrangian using a systematic gradient expansion, and showed how to extend the approximation beyond the lowest order. Here we present a simple derivation, restricting to the lowest order and aiming for conceptual clarity rather than mathematical precision.

Consider non-relativistic motion of a particle with mass $m$ and charge $e$ in a static magnetic field in flat three-dimensional space. The configuration coordinates are the spatial coordinates $x^i$ ($i = 1, 2, 3$), and the Lagrangian $L(x^i, \mathrm{d}x^i/\mathrm{d}t)$ is

$$L = \frac{1}{2}m\mathbf{v}^2 + eA_i\mathbf{v}^i, \qquad (2.1)$$

where $\mathbf{v}^i \equiv \mathrm{d}x^i/\mathrm{d}t$, $\mathbf{v}^2 \equiv g_{ij}\mathbf{v}^i\mathbf{v}^j \equiv \mathbf{v} \cdot \mathbf{v}$, $g_{ij}$ is the (Euclidean) spatial metric, $A_i$ is the magnetic vector potential, which is assumed to depend on $x^k$ but not $t$, and repeated indices are implicitly summed over.[1] The velocity can be decomposed into components parallel and perpendicular to the magnetic field. The latter component is dominated by the gyro motion, but may also have a drift part:

$$\mathbf{v} = \mathbf{v}_\parallel + \mathbf{v}_\text{gyro} + \mathbf{v}_\text{drift}. \qquad (2.2)$$

Accordingly, the kinetic energy decomposes as

$$\frac{1}{2}m\mathbf{v}^2 = \frac{1}{2}m\left(\mathbf{v}_\parallel^2 + \mathbf{v}_\text{gyro}^2 + \mathbf{v}_\text{drift}^2 + 2\mathbf{v}_\text{gyro} \cdot \mathbf{v}_\text{drift}\right). \qquad (2.3)$$

To arrive at the leading-order effective theory, one makes the assumption (to be justified below) that the contribution of the $\mathbf{v}_\text{drift}^2$ term to the equations of motion is negligible so can be dropped, and that $\mathbf{v}_\text{gyro} \cdot \mathbf{v}_\text{drift}$ averages to zero over a gyro period which is short compared with the time scale for other changes, so can also be dropped. The gyro contribution to the kinetic energy is proportional to the product of the gyro angular momentum and the magnetic field strength:

$$\frac{1}{2}m\mathbf{v}_\text{gyro}^2 = \frac{1}{2}m\rho^2\omega_B^2 = -\frac{e}{2m}\ell B = \mu B. \qquad (2.4)$$

Here $\rho$ is the gyroradius, $B = |\mathbf{B}|$ is the magnetic field strength, $\omega_B \equiv -eB/m$ is the gyro frequency, $\ell \equiv m\rho^2\omega_B$ is the (signed) angular momentum of the gyro motion about the direction of the magnetic field line and $\mu \equiv -e\ell/2m$ is the absolute value of the orbital magnetic moment.[2] Crucial to the guiding-centre approximation is the fact that $\ell$ is an adiabatic invariant. It is approximately conserved when the field at the location of the particle changes slowly with respect to the gyro period, so at leading order can be treated as a constant in the Lagrangian.[3] The gyro kinetic energy is thus, in effect, a potential energy, so its negative should be included in the effective Lagrangian as a potential energy term, $-\mu B$.[4]

---

[1]I use here standard index notation for tensors, which distinguishes between contravariant (superscript) and covariant (subscript) indices, and the bold Roman font denotes contravariant vectors without their index. The distinction between co- and contravariant indices will be important when considering quasi-symmetry below, or in applications with non-Cartesian coordinates or curved space–time.

At leading order in the guiding-centre approximation, the motion of the guiding centre is thus governed by the effective Lagrangian:

$$L^{\text{eff}} = \frac{1}{2} m (b_i v^i)^2 + e A_i v^i - \mu B, \qquad (2.5)$$

where

$$b_i = g_{ij} b^j, \qquad b^i = B^i / B. \qquad (2.6)$$

A peculiar feature of the effective Lagrangian (2.5) is that the kinetic energy contribution depends only on the parallel component of the velocity, so the perpendicular component appears only linearly. Therefore no time derivative of the drift velocity appears in the equation of motion, and instead the drift velocity is determined algebraically.

## 3. Equations of motion

The Euler–Lagrange equations for the Lagrangian (2.5) are

$$m \frac{d}{dt}(b_i b_j v^j) = \frac{1}{2} m (b_j b_k)_{,i} v^j v^k + e F_{ij} v^j - \mu B_{,i}, \qquad (3.1)$$

where the comma subscript index $(,i)$ denotes partial derivative with respect to the spatial coordinate $x^i$ and $F_{ij} \equiv A_{j,i} - A_{i,j} = \epsilon_{ijk} B^k$ is the magnetic field strength tensor. With the velocity split into parallel and perpendicular components, denoted as

$$v^i =: v_\parallel b^i + v_\perp^i \qquad (3.2)$$

(where $b_i v_\perp^i = 0$), (3.1) takes the form

$$m \dot{v}_\parallel b_i = -m v_\parallel^2 \kappa_i + 2 m v_\parallel b_{[j,i]} v_\perp^j + e F_{ij} v_\perp^j - \mu B_{,i}, \qquad (3.3)$$

where

$$\kappa_i := 2 b_{[i,j]} b^j \qquad (3.4)$$

---

[2] Ratio $e/2m$ is the gyromagnetic ratio of a point charge. The direction of $e\vec{\ell}$ is antiparallel to $\vec{B}$ for both signs of $e$, so $-e\ell > 0$ for both signs of $e$.

[3] The adiabatic invariance of $\ell$ is a special case of the general theory of adiabatic invariants of phase-space flows, which derives from the exact Poincaré invariant $\oint p_i \, dq^i$, and is equivalent to the conservation of magnetic flux through the gyro orbit.

[4] To formally derive this from the full action, one must account for the fact that $\ell$ is held fixed when the effective action is varied, whereas in the full action $\ell$ varies when the gyroradius and phase vary. To do so one can use the phase-space form of the action for the gryo phase and its conjugate momentum, holding the latter fixed.

is the curvature of the magnetic field lines and $b_{[j,i]} \equiv \frac{1}{2}(b_{j,i} - b_{i,j})$.[5] The term involving $b_{[j,i]}v_\perp^j$ is second order in the background field gradients, since the drift velocity $v_\perp^j$ is first order in those gradients as will be seen momentarily. This term may therefore be dropped in the leading-order approximation.

The parallel component of (3.3), dropping the second order term, is

$$m\dot{v}_\| = -\mu b^i B_{,i}, \quad (3.5)$$

which determines the parallel component of the acceleration. The right-hand side is sometimes called the mirror force, since it may cause reversal of the direction of motion if the increase of 'potential energy' $\mu B$ on a trajectory exceeds the initial parallel kinetic energy. The perpendicular component of (3.3), again dropping the second-order term, is

$$eF_{ij}v_\perp^j = mv_\|^2 \kappa_i + \mu(\nabla_\perp B)_i. \quad (3.6)$$

This is a two-dimensional linear equation for $v_\perp^j$, since $v_\perp^j$ and all terms are orthogonal to $b^i$. To solve it, note that $F_{ij} = B\epsilon_{ij}$, where $\epsilon_{ij} = \epsilon_{ijk}b^k$ is the two-dimensional area form orthogonal to $b^i$, hence

$$v_\perp^k = (eB)^{-1}\epsilon^{ik}\left(mv_\|^2 \kappa_i + \mu B_{,i}\right), \quad (3.7)$$

where $\epsilon^{ik} = \epsilon^{ikl}b_l$ is the inverse of $\epsilon_{ik}$, i.e. $\epsilon^{ik}\epsilon_{jk} = \delta^i{}_j - b^i b_j$. In vector algebra language, the drift velocity (3.7) is $(eB^2)^{-1}$ times the cross product of the magnetic field with the vector obtained by contracting the right-hand side of (3.6) with the inverse metric.

The $\kappa_i$ term in (3.7) is called the curvature drift. The Lorentz force contribution arising from this component of velocity provides the centripetal force that accounts for the curvature of the guiding-centre path following the field line. The $B_{,i}$ term is called the gradient drift. It arises because the radius of curvature of the gyro orbit is larger on the side of the orbit where $B$ is smaller, so on each gyro orbit the particle travels farther in one direction perpendicular to the $B$ gradient than in the opposite direction.

We end this derivation with some comments on the equations of motion:

- The drift velocity is first order in derivatives of the background fields, which justifies the above statement that the term involving $b_{[j,i]}v_\perp^j$ in (3.3) is second order in the background field gradients and can therefore be dropped.[6] It also justifies the dropping of the $v_{\text{drift}}^2$ term in the effective Lagrangian, since that contributes $\sim \mathbf{v} \cdot \nabla \mathbf{v}_{\text{drift}}$ to the equations of motion, which is also second order in the background field gradients. Note that, in addition to terms quadratic in gradients, this includes second gradient terms, both of which must be suppressed if the first-order guiding-centre approximation is to be accurate.[7] This gradient suppression is key to the validity of the guiding-centre approximation.

---

[5]In Cartesian coordinates for Euclidean space, the metric components are constant, so $b_{j,i}b^j = \frac{1}{2}(g_{jk}b^jb^k)_{,i} = \frac{1}{2}(1)_{,i} = 0$, hence the curvature (3.4) is simply $\kappa_i := b_{i,j}b^j$. In an arbitrary coordinate system it can also be written as $= b_{i;j}b^j$, where the semicolon denotes the covariant derivative.

[6]It may nevertheless be useful to preserve this term, so that the solutions would possess the exactly conserved Hamiltonian energy $\frac{1}{2}m(b_iv^i)^2 + \mu B$ that follows from the Lagrangian (2.5).

- The first term in (3.7) is of order $v_\parallel(v_\parallel/\omega_B)(\nabla B/B)$, so is suppressed relative to the parallel velocity by the ratio of the parallel distance travelled in a gyro period to the field gradient length scale. The second term is of order $v_{\text{gyro}}\rho_{\text{gyro}}\nabla B/B$, so is suppressed relative to the gyro velocity by the ratio of the gyro radius to the length scale for changes of the field. For a distribution of charges in thermal equilibrium, the parallel and gyro velocities are typically comparable, so there is just one length ratio governing validity of the guiding-centre approximation. If instead these velocities are vastly different, the approximation scheme may need modification.

- To include any potential energy $\Phi$ one need only make the replacement $\mu B \to \mu B + \Phi$. For example, with an electrostatic potential energy $eV$ this produces electric acceleration along the field lines and the drift velocity contribution $-\epsilon^{ik}E_i/B = (\mathbf{E} \times \mathbf{B})^k/B^2$.

## 4. Quasi-symmetry

The concept of quasi-symmetry was originally introduced as a coordinate symmetry of certain magnetic fields, using coordinates adapted to the magnetic field (Boozer 1983), which can lead to a conserved quantity for guiding-centre motion even when the field admits no geometric symmetry, i.e. no symmetry that is a combination of translation and rotation. Early demonstrations of the existence of field configurations that are approximately quasi-symmetric in this sense were in Nuhrenberg & Zille (1988) and Garren & Boozer (1991).

The formulation of quasi-symmetry criteria in Hamiltonian and Lagrangian frameworks, without the use of magnetic coordinates, is a more recent development (Burby, Kallinikos & MacKay 2020; Rodriguez, Helander & Bhattacharjee 2020). From the latter viewpoint, a quasi-symmetry can be defined by a spatial vector field **u** whose flow preserves (to the required order) the background magnetic field structures in the effective Lagrangian (2.5). Since that Lagrangian does not depend on all components of the metric, a quasi-symmetry need not preserve the metric, i.e. it need not be a geometric symmetry. If the Lagrangian admits such a symmetry, then the guiding-centre motion of charged particles possesses a corresponding conserved quantity, which can aid in the confinement of plasma for controlled nuclear fusion.

Quasi-symmetry is helpful in particular for stellarator designs (Helander 2014; Imbert-Gérard *et al.* 2024), and has been engineered into experimental stellarators (Anderson *et al.* 1995; Qian *et al.* 2023). In stellarators, axial symmetry is sacrificed to avoid the need to externally drive a large toroidal current through the plasma, instead producing the poloidal field needed for confinement by a twisting configuration of external coils.[8] As a symmetry of the guiding-centre Lagrangian, quasi-symmetry can hold without restriction on the current implied by Ampère's law, or the plasma pressure that would be required to balance the Lorentz force on the current in equilibrium. But a quasi-symmetric field configuration is of course only useful for confinement if it can be realised in a stable plasma configuration, with internal electron and ion currents that are consistent with the particle dynamics. Some examples that satisfy this requirement with a good approximation to quasi-symmetry can be found in Ku & Boozer (2010) and Wiedman, Buller & Landreman (2024).

---

[7]I thank R. Andrade e Silva for pointing out the need for a condition on the second-order gradients.

[8]A toroidal 'bootstrap' current develops in a stellarator plasma configuration. In the design of Landreman, Buller & Drevlak (2022), for example, it is an order of magnitude smaller than the current in an axisymmetric tokamak with comparable size and magnetic field strength.

### 4.1. *Quasi-symmetry conditions and conserved quantity*

The background field structures in the effective Lagrangian (2.5) are the magnetic field strength $B$, the vector potential $A_i$ and the unit covector $b_i$. The vector field **u** generates an exact symmetry of these structures if the Lie derivatives[9] of $B$ and $b_i$ along **u** are zero, and that of $A_i$ is a gauge transformation:[10]

$$\mathcal{L}_\mathbf{u} B = 0, \tag{4.1}$$

$$\mathcal{L}_\mathbf{u} A_i = \lambda_{,i}, \tag{4.2}$$

$$\mathcal{L}_\mathbf{u} b_i = 0, \tag{4.3}$$

where $\lambda$ is a scalar function of position.[11] In that case motion governed by the effective Lagrangian admits an exactly conserved quantity. If the gauge parameter $\lambda$ in (4.2) vanishes then the conserved quantity is just the **u** component of the conjugate momentum. If however $\lambda \neq 0$, then the Lagrangian is invariant only up to the total time derivative term $e\dot\lambda$, in which case the conserved quantity is given by

$$K_\mathbf{u} := u^i p_i - e\lambda := m(u^i b_i)(v^j b_j) + e(u^i A_i - \lambda), \tag{4.4}$$

where $p_i$ is the canonical momentum conjugate to the position coordinates $x^i$. Although neither $p_i$ nor $\lambda$ is gauge-invariant, the combination appearing in $K_\mathbf{u}$ is invariant: under $A_i \to A_i + \eta_{,i}$, the change of $u^i p_i$ is $eu^i \eta_{,i}$, and $\mathcal{L}_\mathbf{u} A_k$ in (4.2) changes by $\mathcal{L}_\mathbf{u} \eta_{,k} = (u^i \eta_{,i})_{,k}$, so the change of $e\lambda$ is also $eu^i \eta_{,i}$.

The quasi-symmetry conditions (4.1), (4.2) and (4.3) agree with (20), (21), (22) of Theorem IV.2 of Burby *et al.* (2020). Condition (4.2) looks different from (21) but it is actually equivalent. The exterior derivative of (4.2) implies $\mathcal{L}_\mathbf{u} F = 0$, where $F = dA$ is the field strength 2-form, which differs only in notation from [7, (21)].[12]

---

[9]The Lie derivative $\mathcal{L}_\mathbf{u}$ along **u** of a tensor field is the 'convective derivative' along the flow of **u**. It can be defined without reference to coordinates, but more simply by the partial derivative of the tensor components with respect to a coordinate $q$ in a coordinate system 'adapted' to **u**, i.e. in which **u** is the coordinate vector field '$\partial_q$'. (Here $\partial_q$ is the vector, in the given coordinate system, whose $q$ component is 1 and whose other components are zero. Thus $\partial_q \cdot \nabla = \partial/\partial q$, which motivates the notation.) The flow of a vector field **u** is a geometric symmetry if the Lie derivative of the metric vanishes, $\mathcal{L}_\mathbf{u} g_{ij} = 0$, which in adapted coordinates means simply that the components $g_{ij}$ are independent of the coordinate $q$ along the flow. Such a vector field is called a Killing vector. For an introduction to these methods, see, for example, MacKay (2020) or Schutz (1980).

[10]Under a time-independent gauge transformation, $A_i \to A_i + \alpha_{,i}$, the Lagrangian (2.1) changes by the total time derivative $d(e\lambda)/dt$, so the action and hence the equation of motion is gauge-invariant.

[11]It is essential here that the covariant vector character of $A_i$ and $b_i$, as they appear in the Lagrangian, be respected. Raising their indices by contraction with the inverse metric would modify the symmetry criterion if **u** is not a Killing vector.

[12]A succinct introduction to differential forms is given in Appendix A of Gralla & Jacobson (2014). For a more thorough introduction, see MacKay (2020) or Schutz (1980).

Also, the conserved quantity (4.4) coincides with that defined in (41) of Burby *et al.* (2020): one has $\mathcal{L}_\mathbf{u} A = \mathbf{u} \cdot F + d(\mathbf{u} \cdot A)$,[13] so (4.2) is equivalent to

$$\mathbf{u} \cdot F = d\psi, \qquad (4.5)$$

with $\psi = \lambda - \mathbf{u} \cdot A$. It follows that $K_\mathbf{u}$ may equivalently be written as

$$K_\mathbf{u} = m(u^i b_i)(v^j b_j) - e\psi, \qquad (4.6)$$

with $\psi$ defined by (4.5), as in Burby *et al.* (2020). Expressed as (4.6), each term of $K_\mathbf{u}$ is separately gauge-invariant.

The conservation of $K_\mathbf{u}$ is of little use for the control of charged particle dynamics unless $\psi$ is globally defined, at least within the spatial domain of interest. One can always choose a gauge in which $A$ is globally defined, but global existence of $\lambda$ is not guaranteed in general. Put differently, the symmetry requires $\mathcal{L}_\mathbf{u} F = d(\mathbf{u} \cdot F) = 0$, which (by the Poincaré lemma) implies that $\mathbf{u} \cdot F = d\psi$ for some $\psi$, at least locally. If a global $\psi$ exists, then surfaces of constant $\psi$, called 'flux surfaces', exist. As explained below, the conservation law implies that particle trajectories are approximately confined to such surfaces.[14]

### 4.2. *Existence and role of flux surfaces*

A magnetic field is described by a 2-form $F$ that is closed, $dF = 0$, corresponding to the vector calculus condition $\nabla \cdot \mathbf{B} = 0$. Any closed 2-form in three-dimensional space can be expressed, at least locally, as the wedge product of two exact 1-forms, $F = d\alpha \wedge d\beta$, where $\alpha$ and $\beta$ are functions called Euler potentials.[15] The magnetic field vector is tangent to the surfaces of constant $\alpha$ or $\beta$, i.e. $\mathbf{B} \cdot d\alpha = \mathbf{B} \cdot d\beta = 0$. In terms of $F$, this follows from the antisymmetry of the wedge product, $d\alpha \wedge F = d\alpha \wedge d\alpha \wedge d\beta = 0$, and similarly for $\beta$. The field lines are intersections of these two surfaces. In this sense $\alpha$ and $\beta$ are (local) 'flux functions' for the field. More generally, if $F = d\alpha \wedge \omega$ for some 1-form $\omega$, then $\alpha$ is a flux function for $F$. The function $\psi$ introduced for the quasi-symmetric field in (4.5) is a flux function: $\mathbf{B} \cdot d\psi = \mathbf{B} \cdot u \cdot F = 0$, since for any magnetic field we have $\mathbf{B} \cdot F = \epsilon_{ijk} B^j B^k = 0$.[16] If $\psi$ is globally defined as a single-valued function, and $d\psi$ (4.5) is nowhere zero, level sets of $\psi$ are flux surfaces.

It was argued in Rodriguez *et al.* (2020) that quasi-symmetry implies the existence of flux surfaces. However, that argument assumed implicitly that not only the field strength $F$ but also the vector potential $A$ itself is invariant, i.e. that $\mathcal{L}_\mathbf{u} A = 0$. With this assumption, $\lambda$ in (4.2) vanishes, in which case $\psi = -\mathbf{u} \cdot A$ is indeed globally defined. But does such a gauge exist? A gauge-independent argument was given in Burby *et al.* (2020), invoking the fact that in a domain $Q$ of solid torus topology a circulating magnetic field 'has a closed trajectory realising $H_1(Q)$ [the first homology group], so $\psi$ is global'. The idea here is that $\psi$ is constant along the field lines, and the Brouwer fixed point theorem guarantees the existence of at least one field line that closes after one circuit of the torus, so $\psi$ is continuous on at least one loop around the torus, which is enough to imply that it is globally defined. A different argument was given in § 1.2.3 of Rodriguez (2022), formulated using vector calculus. For the case of a solid torus region, and in the language of differential forms, that argument can be presented as follows.

---

[13] I use the notation $\mathbf{u} \cdot$ for the interior product $i_\mathbf{u}$, i.e. contraction of the vector $\mathbf{u}$ with the first slot in the differential form to its right. Here and below, use is made of 'Cartan's magic formula' for the action of the Lie derivative on a differential form of any rank, $\mathcal{L}_\mathbf{u} = i_\mathbf{u} d + d i_\mathbf{u}$.

[14] Flux surfaces must be toroidal if they are compact without boundary and the magnetic field is everywhere non-zero (Boozer 2004).

[15] This is also the case for ideal magnetohydrodynamic or force-free electromagnetic fields in four space–time dimensions, since then $F$ is locally the wedge product of 2-forms. For a proof, see § 3.2 of Gralla & Jacobson (2014).

[16] To determine $\omega$ such that $F = d\psi \wedge \omega$, note that $F \wedge F$ is a 4-form, so it vanishes in three dimensions, hence $0 = \mathbf{v} \cdot \mathbf{w} \cdot (F \wedge F) = 2(\mathbf{v} \cdot \mathbf{w} \cdot F)F + 2(\mathbf{v} \cdot F) \wedge (\mathbf{w} \cdot F)$ (using the fact that interior product with a vector is an anti-derivation on the algebra of differential forms). If $\mathbf{v}$ and $\mathbf{w}$ are any two vectors with $\mathbf{v} \cdot \mathbf{w} \cdot F \neq 0$, we thus have $F = (\mathbf{v} \cdot F) \wedge (\mathbf{w} \cdot F)/\mathbf{w} \cdot \mathbf{v} \cdot F$. Taking $\mathbf{v} = \mathbf{u}$, this yields $F = d\psi \wedge (\mathbf{w} \cdot F)/(\mathbf{w} \cdot d\psi)$.

Suppose that in a solid torus domain we have $\mathbf{u} \cdot F = d\psi$, with $\mathbf{u} \cdot F$ everywhere differentiable and non-vanishing. The fact that toroidal loops are non-contractible could in general present an obstruction to the existence of a global flux function $\psi$, because $\psi$ could be like the angle on a circle, which has a discontinuity. Suppose $\psi$ has a jump $\Delta\psi$ across some poloidal cross-section of the torus. Call that cross-section the 'cut'. Then $\Delta\psi$ must be independent of position on the cut, since $\mathbf{u} \cdot F$, and hence $d\psi$, is continuous, so the change of $\psi$ along a curve tangent to the cut must be the same on both sides of the cut. The solid torus minus the cut is contractible, and in that contractible domain we have (using $dF = 0$) $d(\psi F) = d\psi \wedge F = d\psi \wedge d\psi \wedge \omega = 0$. The integral of $d(\psi F)$ over the cut torus domain thus vanishes, and by the Stokes theorem it is also equal to $\Delta\psi$ times the magnetic flux $\int_{\text{cut}} F$ through the cut. As long as that flux does not vanish, it follows that $\Delta\psi = 0$. That is, $\psi$ must in fact be continuous, and hence globally defined.[17]

Although the conserved quantity (4.6) has two terms, it is typically dominated by $\psi$ in the plasma confinement setting, in the sense that the conservation law approximately confines the charge to surfaces of constant $\psi$. To estimate the distance $s$ of variations away from a constant $\psi$ surface, note that the change of $\psi$ over a displacement $\mathbf{s}$ is

$$\delta\psi \sim \mathbf{s} \cdot d\psi = \mathbf{s} \cdot \mathbf{u} \cdot F = \mathbf{s} \cdot (\mathbf{u} \times \mathbf{B}) \sim s u_\perp B. \quad (4.7)$$

For $e\,\delta\psi$ to cancel variations of the $mu_\parallel v_\parallel$ term in (4.6), $s$ must therefore be of order (cf. Rodriguez *et al.* 2020)

$$s \sim \frac{u_\parallel}{u_\perp} \frac{v_\parallel}{\omega_B} = \frac{u_\parallel}{u_\perp} \frac{v_\parallel}{v_{\text{gyro}}} \rho_{\text{gyro}}. \quad (4.8)$$

If the ratios in the last term of (4.8) are of order unity, the deviation from a constant $\psi$ surface is of the order of the gyro radius.

### 4.3. *Weak quasi-symmetry*

It was pointed out in Rodriguez *et al.* (2020) that the last of the symmetry conditions above, (4.3), is too strong. It is indeed required for an exact symmetry of the effective Lagrangian, but that Lagrangian is only derived at leading order in the background field derivative expansion, which is why in the equations of motion we dropped the second-order term in (3.3). For deriving a conserved quantity at leading order, symmetry of the Lagrangian should therefore be required only at leading order. In particular, this means that one should require not that $\mathcal{L}_\mathbf{u} b_i$ vanish, but only that $v^i \mathcal{L}_\mathbf{u} b_i = v_\parallel b^i \mathcal{L}_\mathbf{u} b_i + v^i_\perp \mathcal{L}_\mathbf{u} b_i$ vanish at leading order (where $v^i$ is split as in (3.2)). The drift velocity $v^i_\perp$ is a first-order quantity, and $\mathcal{L}_\mathbf{u} b_i$ is also first order in derivatives. Hence the appropriate first-order symmetry condition is not the three conditions (4.3) but rather the single condition that the component of $\mathcal{L}_\mathbf{u} b_i$ along $b^i$ vanish:

$$b^i \mathcal{L}_\mathbf{u} b_i = 0. \tag{4.9}$$

---

[17] For an (artificial) example where the flux through the cut does vanish, consider a magnetic monopole line charge. In cylindrical coordinates this has field strength $F = dz \wedge d\varphi$. This admits the symmetry vector field $u = \partial_z$, and $u \cdot F = d\varphi$, so the local flux function is $\psi = \varphi$, which is not globally defined.

Taken together with (4.1) and (4.2), the additional condition (4.9) holds if and only if $\mathcal{L}_\mathbf{u} \epsilon^{ijk} = \epsilon^{ijk} \nabla \cdot \mathbf{u} = 0$, i.e. if the flow of $\mathbf{u}$ is volume-preserving (Rodriguez *et al.* 2020):

$$0 = \mathcal{L}_\mathbf{u} B = \mathcal{L}_\mathbf{u}(b_i B^i) = b_i \mathcal{L}_\mathbf{u} B^i = \frac{1}{2} b_i \mathcal{L}_\mathbf{u}(\epsilon^{ijk} F_{jk}) = \frac{1}{2} b_i F_{jk} \mathcal{L}_\mathbf{u} \epsilon^{ijk} = B \nabla \cdot \mathbf{u}. \tag{4.10}$$

The collection of first-order quasi-symmetry constraints can thus be presented as

$$\mathcal{L}_\mathbf{u} B = 0, \tag{4.11}$$

$$\mathcal{L}_\mathbf{u} F = 0, \tag{4.12}$$

$$\mathcal{L}_\mathbf{u} \epsilon = 0, \tag{4.13}$$

where $\epsilon$ is the volume 3-form.[18] In Rodriguez *et al.* (2020) this collection of constraints was dubbed 'weak quasi-symmetry', to distinguish it from the stronger version (4.1), (4.2), (4.3) which was dubbed 'strong quasi-symmetry'. It is worth emphasising that conditions (4.12) and (4.13) together imply that the contravariant magnetic field vector $B^i$ is invariant, $\mathcal{L}_\mathbf{u} B^i = \frac{1}{2} \mathcal{L}_\mathbf{u}(\epsilon^{ijk} F_{jk}) = 0$, and (4.11) then implies that also $b^i$ is invariant, but the covariant vectors $B_i$ and $b_i$ are not invariant.

Returning for a moment to strong quasi-symmetry, the condition (4.3) can be expressed as $\mathcal{L}_\mathbf{u} b_i = \mathcal{L}_\mathbf{u}((*F)_i/B) = 0$, where $*F$ ($\equiv B_i$) is the three-dimensional Hodge dual of the field strength.[19] Given the constancy of $B$ (4.1), the constancy of $b_i$ (4.3) may thus be replaced by $\mathcal{L}_\mathbf{u} *F = 0$. Together with $\mathcal{L}_\mathbf{u} F = 0$ (which is equivalent to (4.2)) this implies that $\mathcal{L}_\mathbf{u}(F \wedge *F) = \mathcal{L}_\mathbf{u}(B^2 \epsilon) = 0$. Condition (4.1) then implies that $\mathcal{L}_\mathbf{u} \epsilon = 0$. Alternatively, we may trade the condition (4.1) for $\mathcal{L}_\mathbf{u} \epsilon = 0$, so that the strong quasi-symmetry conditions become

$$\mathcal{L}_\mathbf{u} *F = 0, \tag{4.14}$$

$$\mathcal{L}_\mathbf{u} F = 0, \tag{4.15}$$

$$\mathcal{L}_\mathbf{u} \epsilon = 0. \tag{4.16}$$

---

[18] $\mathcal{L}_\mathbf{u} \epsilon^{ijk} = 0$ implies that also the Lie derivative of the covariant volume form $\epsilon_{ijk}$ vanishes: $\epsilon^{ijk} \mathcal{L}_\mathbf{u} \epsilon_{ijk} = \mathcal{L}_\mathbf{u}(\epsilon^{ijk} \epsilon_{ijk}) = \mathcal{L}_\mathbf{u}(6) = 0 \implies \mathcal{L}_\mathbf{u} \epsilon_{ijk} = 0$ (since the space of totally antisymmetric rank-3 tensors in three dimensions is one-dimensional).

[19] In tensor notation, $*F_i \equiv \frac{1}{2} g_{il} \epsilon^{jkl} F_{jk}$. An often more practical definition is that the dual of a $p$-form $C$ in an $n$-dimensional space is the orthogonal $(n-p)$-form of the same norm, with sign chosen so that $C \wedge *C = \langle C, C \rangle \epsilon$, where $\epsilon$ is the volume element and $\langle C, C \rangle := (1/p!) C^{i...j} C_{i...j}$ is the squared norm of $C$. In the present example, $F \wedge *F = B^2 \epsilon$. The Hodge dual is defined relative to a choice of orientation, i.e. a sign of the volume form.

In this formulation, strong thus differs from weak quasi-symmetry only in that symmetry of $B$ is replaced by symmetry of $*F$.

Although a quasi-symmetry is generally not a metric isometry, it must preserve some aspects of the metric. Condition (4.13) states that a weak quasi-symmetry preserves the volume element; and, since $\mathcal{L}_\mathbf{u} b^i = 0$, the projection of the Lie derivative of the metric onto the magnetic field line direction must vanish:

$$b^i b^j \mathcal{L}_\mathbf{u} g_{ij} = \mathcal{L}_\mathbf{u}(b^i b^j g_{ij}) = \mathcal{L}_\mathbf{u} 1 = 0. \qquad (4.17)$$

This still allows the metric in the directions perpendicular to the magnetic field to be deformed by the flow. However, that deformation must preserve the area element orthogonal to the field lines:

$$\mathcal{L}_\mathbf{u} \epsilon_{ij} = \mathcal{L}_\mathbf{u}(\epsilon_{ijk} b^k) = 0. \qquad (4.18)$$

Since $\mathcal{L}_\mathbf{u} b_i = \mathcal{L}_\mathbf{u}(b^j g_{ij}) = b^j \mathcal{L}_\mathbf{u} g_{ij}$, a strong quasi-symmetry must preserve even more of the metric, $b^j \mathcal{L}_\mathbf{u} g_{ij} = 0$. Note that in the special case where a quasi-symmetry is a metric isometry, it is necessarily a strong quasi-symmetry, since then $\mathcal{L}_\mathbf{u} b^i = 0$ implies $\mathcal{L}_\mathbf{u} b_i = 0$.

### 4.4. *Realizations and relations between quasi-symmetry types*

Non-isometric examples of configurations with exact weak quasi-symmetry in a topologically toroidal domain were found in Sato (2022). These are not useful in practice, since they would require anisotropic material pressure for force balance on the electric current and lack the field line twist ('rotational transform') required for confinement of the plasma. Nevertheless, they provide a helpful example, which will be discussed for illustrative purposes in the next section.

If additional conditions, such as restriction to vacuum fields or magnetohydrodynamic equilibrium with isotropic material pressure, are imposed, it might be that a field configuration can be quasi-symmetric only if it is invariant under a geometric isometry (Garren & Boozer 1991), in which case stellarator fields could never be exactly quasi-symmetric. That said, vacuum configurations that admit no isometry but are very nearly quasi-symmetric have been found (Landreman & Paul 2022). These field configurations have been found by numerical optimisation methods, and do not so far appear to converge to exact quasi-symmetry. Their quasi-symmetry violations scale with the cube of the aspect ratio (Landreman & Sengupta 2019), as predicted by the analysis of Garren & Boozer (1991), but the numerical coefficient of the violation parameter is extremely small, of order $10^{-3}$.

Existing work involving the construction of quasi-symmetric configurations by optimisation has used the Boozer (1983) definition of quasi-symmetry, which implies strong quasi-symmetry. In fact, that definition is equivalent to both strong and weak quasi-symmetry (apart from exceptional cases) under the assumption that nested toroidal flux surfaces exist with no current flow across them (the current flow restriction is implied by magnetohydrodynamic equilibrium with isotropic pressure that is constant on flux surfaces) (Simakov & Helander 2011). The equivalence with weak quasi-symmetry can be established via the triple product characterisation of the latter, $\nabla \psi \times \nabla B \cdot \nabla (\mathbf{B} \cdot \nabla B) = 0$ (Rodriguez *et al.* 2020), equivalently, $\mathrm{d}\psi \wedge \mathrm{d}B \wedge \mathrm{d}|F \wedge \mathrm{d}B| = 0$.

## 5. Examples

Two examples will serve here to illustrate drift velocity and quasi-symmetry, and computational techniques with differential forms.

The first example is the magnetic field of an infinite line current $I$ along the $z$ axis, $\mathbf{B} = (\mu_0 I/2\pi r)\hat{\varphi}$ in cylindrical coordinates, with $I > 0$. A positively charged particle that gyrates around a field line will have a drift velocity (3.7) in the $\hat{z}$ direction, since both the gradient of $B$ and the curvature $\kappa^i$ (3.4) of the circular field line point in the $-\hat{r}$ direction. The gradient contribution is $\mu/er = v_{\text{gyro}}(\rho_{\text{gyro}}/2r)$, and the curvature contribution is $mv_\parallel^2/eBr = v_\parallel(v_\parallel/v_{\text{gyro}})(\rho_{\text{gyro}}/r)$, so their ratio is $\frac{1}{2}(v_{\text{gyro}}/v_\parallel)^2$.

The field strength 2-form corresponding to the line current field is

$$F = dz \wedge d\alpha, \tag{5.1}$$

with $\alpha = \alpha(r) = (\mu_0 I/2\pi)\ln r$. More generally, (5.1) is solenoidal ($dF = 0$) with any function $\alpha(r)$ (since $d$ obeys the Liebniz rule with respect to wedge products, and $d^2 = 0$). Any function of $z$ and $r$ is a flux function for this field, and it has $z$-translation and $\varphi$-rotation symmetry. It also admits a wider class of quasi-symmetries, generated by the vector field

$$\mathbf{u} = \beta(r)\partial_z + \gamma(r)\partial_\varphi, \tag{5.2}$$

where $\partial_z$ and $\partial_\varphi$ are cylindrical coordinate vector fields (cf. the first footnote of § 4) and $\beta(r)$ and $\gamma(r)$ are arbitrary functions. This is also a strong quasi-symmetry if $\gamma$ is constant, and is a metric isometry only when both $\beta$ and $\gamma$ are constants.

Let us verify the quasi-symmetry conditions:

- (4.11): $B = |\alpha'|$ is independent of $z$ and $\varphi$, hence $\mathcal{L}_\mathbf{u} B = 0$.
- (4.12): $\mathcal{L}_\mathbf{u} F = d(\mathbf{u} \cdot F) = d(\beta\, d\alpha) = d(\beta\alpha'\, dr) = (\beta\alpha')'\, dr \wedge dr = 0$.
- (4.13): $\nabla \cdot \mathbf{u} = 0$ since $\mathbf{u}$ has only $z$ and $\varphi$ components yet no $z$ or $\varphi$ dependence.

As for the strong quasi-symmetry condition (4.16), for the field (5.1) we have

$$*F = \alpha' *(dz \wedge dr) = \alpha' r\, d\varphi \tag{5.3}$$

(where we have chosen the orientation $\epsilon = dz \wedge dr \wedge r\, d\varphi$). Since $\mathcal{L}_\mathbf{u} r = 0$, we need only evaluate $\mathcal{L}_\mathbf{u} d\varphi = d\mathcal{L}_\mathbf{u} \varphi = d\gamma$. Thus $\mathbf{u}$ generates a strong quasi-symmetry for the field (5.1) if and only if $\gamma$ is constant, and that is a metric isometry only if also $\beta$ is constant.

Next consider the less symmetric, but still analytically tractable, example found in Sato (2022). It consists of the field of an infinite line current, deformed by components in the $\hat{\varphi}$ and $\hat{z}$ directions, but I will generalise that slightly here. The field strength 2-form is

$$F = d(z - h) \wedge d\alpha, \tag{5.4}$$

with

$$h = h(\eta) = h(\eta + 2\pi), \qquad \eta \equiv n\varphi + q(r)z + \nu(r), \tag{5.5}$$

where $n$ is an integer, $h(\eta)$ is an arbitrary periodic function of $\eta$ and $\alpha(r)$, $q(r)$ and $\nu(r)$ are arbitrary functions. For this field, any function of $z - h$ and $r$ is a flux function. For illustration Sato chooses the toroidal boundary of the domain to be a level set of the flux function

$$\Psi = \frac{1}{2}\left[(r-r_0)^2 + \mathcal{E}(z-h)^2\right], \tag{5.6}$$

where $r_0$ and $\mathcal{E}$ are positive constants. The form of the quasi-symmetry vector field is

$$\mathbf{u} = \sigma(n\partial_z - q\partial_\varphi), \tag{5.7}$$

where $\sigma = \sigma(r)$ is an arbitrary function. That is, the effective Lagrangian (2.5), with the field (5.4), is invariant at first order under the $r$-dependent combination (5.7) of $z$ translation and $\varphi$ rotation, which is not an isometry of the metric. The 'strong quasi-symmetry' condition (4.3) does not hold, unless either (5.4) reduces to the previous case (5.1), or $d\sigma = dq = 0$ in which case (5.7) is a Killing vector and (5.4) has helical geometric symmetry.

We now verify the quasi-symmetry conditions for (5.4) with (5.7):

- (4.13): $\mathbf{u}$ has only $z$ and $\varphi$ components, yet no $z$ or $\varphi$ dependence, so $\nabla \cdot \mathbf{u} = 0$.
- (4.12): $\mathcal{L}_\mathbf{u} r = 0$ and $\mathcal{L}_\mathbf{u} \eta = 0$, so $\mathcal{L}_\mathbf{u}\alpha = \mathcal{L}_\mathbf{u} h = 0$; and $\mathcal{L}_\mathbf{u} z = n\sigma$, so $\mathcal{L}_\mathbf{u} dz = n\, d\sigma$. Hence $\mathcal{L}_\mathbf{u} F = n\, d\sigma \wedge d\alpha = n\sigma'\alpha'\, dr \wedge dr = 0$.
- (4.11): (5.4) can be written as

$$F = (\alpha' - h'q)\, dz \wedge dr - nh'\, d\varphi \wedge dr. \tag{5.8}$$

The differential forms $dr$, $dz$ and $r\,d\varphi$ are orthonormal, so $B^2 = (\alpha' - h'q)^2 + (nh'/r)^2$. Since this is a function of only $r$ and $\eta$, (4.11) follows.

As for the strong quasi-symmetry condition (4.16), for the field (5.8) we have

$$*F = (\alpha' - h'q)r\, d\varphi - (nh'/r)\, dz. \tag{5.9}$$

Since $\mathcal{L}_\mathbf{u} r = \mathcal{L}_\mathbf{u} \eta = 0$, $\mathcal{L}_\mathbf{u} d\varphi = -d(\sigma q)$ and $\mathcal{L}_\mathbf{u} dz = n\, d\sigma$, we have

$$\mathcal{L}_\mathbf{u} * F = -(\alpha' - h'q)r\, d(\sigma q) - (n^2 h'/r)\, d\sigma. \tag{5.10}$$

If this is to vanish for all $r$ and $\eta$, the terms involving $h$ must vanish themselves, hence the term not involving $h$ must also vanish by itself. Since $\alpha' \neq 0$ (otherwise $F = 0$) this requires $d(\sigma q) = 0$, and therefore either $n = 0$ (axisymmetry), or $h' = 0$ (which is equivalent to $h = 0$) or $d\sigma = 0$. The only interesting case is the last one, which together with $d(\sigma q) = 0$ implies that also $dq = 0$. In that case, $\mathbf{u}$ becomes a constant linear combination of $z$ translation and $\varphi$ rotation: it is a helical metric isometry.

The approximately conserved quantity $K_\mathbf{u}$ conjugate to the quasi-symmetry (5.7) of the field (5.4) can be computed from (4.6). We have $\mathbf{u} \cdot F = (\mathbf{u} \cdot dz)d\alpha = n\sigma\, d\alpha =: d\psi$, where the existence of $\psi$ in the last step follows since both $\sigma$ and $\alpha$ are functions only of $r$. The approximately conserved quantity is thus approximately just the $r$ coordinate.

## 6. Conclusion

Our aim was to provide a concise and instructive introduction to interesting well-known results, and to cast a novel light on the subject through the use of less familiar mathematical techniques. Tensor analysis provides manifestly coordinate independent equations which can be evaluated in any convenient coordinate system with the help of the metric components, and exterior calculus provides a streamlined simplification of many calculations.

We have shown how to obtain an effective Lagrangian (2.5) governing dynamics in the guiding-centre approximation, at first order in the magnetic field gradient, by a simple reasoning involving coarse-graining over the gyro motion and invoking the adiabatic invariance of the gyromagnetic moment. We have also shown how the resulting Euler–Lagrange equations algebraically determine a drift velocity (3.7) perpendicular to the magnetic field lines, on account of the absence of transverse kinetic energy in the Lagrangian.

Since the effective Lagrangian (2.5) does not involve the full Euclidean spatial metric in the kinetic energy term, the action can admit a 'quasi-symmetry', without admitting a metric isometry. Since the effective Lagrangian is accurate only to first order in the field gradients, the quasi-symmetry need only hold to that order to produce a quantity conserved at that order. We have explained how these considerations give rise to what has been called 'weak quasi-symmtetry'. The use of the Lie derivative and differential forms has provided a particularly natural language for the derivation and structure of the results. An explicit example of a quasi-symmetric field and its conserved quantity, taken from the literature, has been used to illustrate these ideas and mathematical techniques.

### Acknowledgements

I am grateful to M. Landreman for extensive instruction and guidance to the literature, to I. Rothstein for discussions about the effective Lagrangian description of guiding-centre motion and to both of them for helpful comments and suggestions on drafts of this paper. I also thank N. Kallinikos and J.W. Burby for explanation of some topological reasoning in their paper, and I am grateful to the anonymous reviewers, whose critique and suggestions led to substantial improvements in the presentation.

### Funding

This work was supported in part by NSF grant PHY-2309634. I am grateful for the hospitality of Perimeter Institute where part of this work was carried out. Research at Perimeter Institute is supported in part by the Government of Canada through the Department of Innovation, Science and Economic Development and by the Province of Ontario through the Ministry of Colleges and Universities.

### Declaration of interest

The author reports no conflict of interest.

### REFERENCES

ALFVÉN, H. 1950 *Cosmical Electrodynamics*. Oxford University Press.
ANDERSON, F.S.B., ALMAGRI, A.F., ANDERSON, D.T., MATTHEWS, P.G., TALMADGE, J.N. & SHOHET, J.L. 1995 The helically symmetric experiment (HSX), goals, design and status. *Fusion Technol.* **27** (3T), 273–277.
BOOZER, A.H. 1983 Transport and isomorphic equilibria. *Phys. Fluids* **26** (2), 496–499.
BOOZER, A.H. 2004 Physics of magnetically confined plasmas. *Rev. Mod. Phys.* **76** (4), 1071–1141.

BURBY, J.W., KALLINIKOS, N. & MACKAY, R.S. 2020 Some mathematics for quasi-symmetry. *J. Math. Phys.* **61** (9), 093503, 1–22.

GARREN, D. & BOOZER, A.H. 1991 Existence of quasihelically symmetric stellarators. *Phys. Fluids B* **3** (10), 2822–2834.

GRALLA, S.E. & JACOBSON, T. 2014 Spacetime approach to force-free magnetospheres. *Mon. Not. R. Astron. Soc.* **445** (3), 2500–2534.

HAZELTINE, R. & WAELBROECK, F. 1998 *The Framework Of Plasma Physics*. Avalon Publishing.

HELANDER, P. 2014 Theory of plasma confinement in non-axisymmetric magnetic fields. *Rep. Prog. Phys.* **77** (8), 087001.

IMBERT-GÉRARD, L., PAUL, E. & WRIGHT, A. 2024 An introduction to stellarators: from magnetic fields to symmetries and optimization. *Soc. Indust. Appl. Maths*. https://doi.org/10.1137/1.9781611978223

KU, L. & BOOZER, A. 2010 New classes of quasi-helically symmetric stellarators. *Nucl. Fusion* **51** (1), 013004.

LANDREMAN, M., BULLER, S. & DREVLAK, M. 2022 Optimization of quasi-symmetric stellarators with self-consistent bootstrap current and energetic particle confinement. *Phys. Plasmas* **29** (8), 082501, 1–16.

LANDREMAN, M. & PAUL, E. 2022 Magnetic fields with precise quasisymmetry for plasma confinement. *Phys. Rev. Lett.* **128** (3), 035001.

LANDREMAN, M. & SENGUPTA, W. 2019 Constructing stellarators with quasisymmetry to high order. *J. Plasma Phys.* **85** (6), 815850601.

LITTLEJOHN, R.G. 1983 Variational principles of guiding centre motion. *J. Plasma Phys.* **29** (1), 111–125.

MACKAY, R.S. 2020 Differential forms for plasma physics. *J. Plasma Phys.* **86** (1), 925860101.

NORTHROP, T. 1963 *The Adiabatic Motion of Charged Particles*. Interscience Publishers.

NÜHRENBERG, J. & ZILLE, R. 1988 Quasi-helically symmetric toroidal stellarators. *Phys. Lett. A* **129** (2), 113–117.

QIAN, T., *et al.* et al. 2023 Design and construction of the MUSE permanent magnet stellarator. *J. Plasma Phys.* **89** (5), 955890502.

RODRIGUEZ, E. 2022 Quasisymmetry *Phd thesis*, Princeton University, Available at https://dataspace.princeton.edu/handle/88435/dsp01x633f4257

RODRIGUEZ, E., HELANDER, P. & BHATTACHARJEE, A. 2020 Necessary and sufficient conditions for quasisymmetry. *Phys. Plasmas* **27** (6), 062501, 1–5.

SATO, N. 2022 Existence of weakly quasisymmetric magnetic fields without rotational transform in asymmetric toroidal domains. *Sci. Rep-UK* **12** (1), 11322.

SCHUTZ, B. 1980 *Geometrical Methods of Mathematical Physics*. Cambridge University Press.

SIMAKOV, A.N. & HELANDER, P. 2011 Plasma rotation in a quasi-symmetric stellarator. *Plasma Phys. Control. Fusion* **53** (2), 024005.

TAYLOR, J. 1964 Equilibrium and stability of plasma in arbitrary mirror fields. *Phys. Fluids* **7** (6), 767–773.

WIEDMAN, A., BULLER, S. & LANDREMAN, M. 2024 Coil optimization for quasi-helically symmetric stellarator configurations. *J. Plasma Phys.* **90** (3), 905900307.